\documentclass[conference]{IEEEtran}
\IEEEoverridecommandlockouts
\usepackage{cite}
\usepackage{amsmath,amssymb,amsfonts}
\usepackage{algorithmic}
\usepackage{graphicx}
\usepackage{textcomp}
\usepackage{xcolor}
\usepackage[tikz]{bclogo}
\usepackage[framemethod=tikz]{mdframed}
\usepackage{lipsum}

\usepackage[many]{tcolorbox}
\usepackage{multirow}
\usepackage{array}
\usepackage{arydshln}
 \usepackage[subtle,tracking=normal]{savetrees}
\definecolor{bgblue}{RGB}{245,243,253}
\definecolor{ttblue}{RGB}{91,194,224}

\mdfdefinestyle{mystyle}{%
  rightline=true,
  innerleftmargin=10,
  innerrightmargin=10,
  outerlinewidth=0pt,
  topline=false,
  rightline=true,
  bottomline=false,
  skipabove=\topsep,
  skipbelow=\topsep
}

\newtcolorbox{myboxi}[1][]{
  breakable,
  title=#1,
  colback=white,
  colbacktitle=white,
  coltitle=black,
  fonttitle=\bfseries,
  bottomrule=0pt,
  toprule=0pt,
  leftrule=3pt,
  rightrule=3pt,
  titlerule=0pt,
  arc=0pt,
  outer arc=0pt,
  colframe=black,
}

\newtcolorbox{myboxii}[1][]{
  breakable,
  freelance,
  title=#1,
  colback=white,
  colbacktitle=white,
  coltitle=black,
  fonttitle=\bfseries,
  bottomrule=0pt,
  boxrule=0pt,
  colframe=white,
  overlay unbroken and first={
  \draw[black,line width=3pt]
    ([xshift=5pt]frame.north west) -- 
    (frame.north west) -- 
    (frame.south west);
  \draw[black,line width=3pt]
    ([xshift=-5pt]frame.north east) -- 
    (frame.north east) -- 
    (frame.south east);
  },
  overlay unbroken app={
  \draw[black,line width=3pt,line cap=rect]
    (frame.south west) -- 
    ([xshift=5pt]frame.south west);
  \draw[black,line width=3pt,line cap=rect]
    (frame.south east) -- 
    ([xshift=-5pt]frame.south east);
  },
  overlay middle and last={
  \draw[black,line width=3pt]
    (frame.north west) -- 
    (frame.south west);
  \draw[black,line width=3pt]
    (frame.north east) -- 
    (frame.south east);
  },
  overlay last app={
  \draw[black,line width=3pt,line cap=rect]
    (frame.south west) --
    ([xshift=5pt]frame.south west);
  \draw[black,line width=3pt,line cap=rect]
    (frame.south east) --
    ([xshift=-5pt]frame.south east);
  },
}
\usepackage{xcolor}
\usepackage{framed}
\colorlet{shadecolor}{lightgray}

\def\BibTeX{{\rm B\kern-.05em{\sc i\kern-.025em b}\kern-.08em
    T\kern-.1667em\lower.7ex\hbox{E}\kern-.125emX}}
\begin{document}

\title{Debiasing Architectural Decision-Making: An Experiment With Students and Practitioners
\thanks{ Part of this work was supported by Funetec-PB – Ct.: Phoebus 01/24 and Snet 04/22.}
}

\author{\IEEEauthorblockN{1\textsuperscript{st} Klara Borowa}
\IEEEauthorblockA{\textit{Warsaw University of Technology,}\\ 
\textit{Institute of Control and Computation Engineering}\\
Warsaw, Poland \\
klara.borowa@pw.edu.pl}
\and
\IEEEauthorblockN{2\textsuperscript{nd} Rodrigo Rebouças de Almeida}
\IEEEauthorblockA{\textit{Federal University of Paraíba} \\\textit{Department of Exact Sciences} \\
Rio Tinto/PB, Brazil \\
rodrigor@dcx.ufpb.br}
\and
\IEEEauthorblockN{3\textsuperscript{rd} Marion Wiese}
\IEEEauthorblockA{\textit{Universität Hamburg} \\
\textit{Department of Informatics}\\
Hamburg, Germany \\
marion.wiese@uni-hamburg.de}
}

\maketitle

\begin{abstract}
Cognitive biases are predictable, systematic errors in human reasoning. They influence decision-making in various areas, including architectural decision-making, where architects face many choices. \color{black} For example, anchoring can cause architects to unconsciously prefer the first architectural solution that they came up with, without considering any solution alternatives. \color{black} Prior research suggests that training individuals in debiasing techniques during a practical workshop can help reduce the impact of biases.
The goal of this study was to design and evaluate a debiasing workshop with individuals at various stages of their professional careers.
To test the workshop's effectiveness, we performed an experiment with 16 students and 20 practitioners, split into control and workshop group pairs. We recorded and analyzed their think-aloud discussions about improving the architectures of systems they collaborated on.
The workshop improved the participants' argumentation when discussing architectural decisions and increased the use of debiasing techniques taught during the workshop. This led to the successful reduction of the researched biases' occurrences. In particular, anchoring and optimism bias occurrences decreased significantly.
We also found that practitioners were more susceptible to cognitive biases than students, so the workshop had a more substantial impact on practitioners.
We assume that the practitioners' attachment to their systems may be the cause of their susceptibility to biases.
\color{black} Finally, we identified factors that may reduce the effectiveness of the debiasing workshop. On that basis, we prepared a set of teaching suggestions for educators. \color{black} 
Overall, we recommend using this workshop to educate both students and experienced practitioners about the typical harmful influences of cognitive bias on architectural decisions and how to avoid them.
\end{abstract}

\begin{IEEEkeywords}
Software Architecture, Architectural Decisions, Architectural Decision-Making, Cognitive bias, Debiasing
\end{IEEEkeywords}

\section{Introduction}
\label{sec:Introduction}
        
        \textbf{Cognitive biases} are systematic errors caused by the heuristics the human mind uses to reduce the complexity of various tasks \cite{Tversky1974}, including decision-making.
        Cognitive biases distort human decision-making in various domains: from clinicians making erroneous medical diagnoses \cite{norman2017causes} to developers copying code without reading it \cite{Chattopadhyay2020}.
        Stacy and Macmillan \cite{Stacy1995} first observed the possible impact of cognitive biases in the realm of software engineering. 
        \color{black} In particular, how \textit{representativeness,} \textit{availability bias} and \textit{confirmation bias} could influence software developers' daily activities. 
        \color{black}
        Since then, research on cognitive biases has significantly expanded to include at least 36 biases and their impact on various software engineering knowledge areas~\cite{Mohanani2018}.
        The influence of cognitive biases is particularly impactful in architectural decision-making (ADM)~\cite{Tang2011}~\cite{VanVliet2016}\color{black}, \color{black} since software architecture can be considered as a set of design decisions \cite{jansen2005software}.
        However, few studies have focused on behavioral factors in ADM, with even fewer containing any empirical validation of decision-making techniques \cite{Razavian2019}.
        Notable examples of studies that have focused on improving the process of ADM include those that assess the usefulness of reflection \color{black}-- \color{black}Razavian et al. \cite{razavian2016two} (experiment on students), and Tang et al. \cite{Tang2018} (experiment on students and professionals).

        \textbf{Debiasing} is a process that improves one's judgment by using techniques decreasing the impact of a particular cognitive bias~\cite{fischhoff1982debiasing}.
        A workshop can effectively mitigate some influences of cognitive biases on software practitioners'~\cite{Shepperd2018}.
        As far as we know, the only empirically tested debiasing training for ADM was conducted on a group of students with limited experience in software development~\cite{Borowa2022}. This study had students perform theoretical tasks, and \color{black} employed a pretest-posttest design. \color{black} 
        Given the doubts expressed by the software engineering community about the validity of the findings in experiments with student participants~\cite{galster2023empirical}~\cite{ baltes2022sampling}, as well as numerous threats to validity associated with pretest-posttest experiments~\cite{Knapp2016}, further evaluation of debiasing through workshops is warranted.
        Therefore, the \textbf{goal} of this study is to conduct a debiasing intervention with a control group to ascertain its effectiveness and investigate its impact on individuals at various stages of their careers, i.e., students and practitioners. 
         
        \noindent
            As such, we strive to answer the following research question:
            \begin{itemize}
                \item [] \textbf{RQ: How does the proposed debiasing architectural decision-making workshop influence students and practitioners?}
            \end{itemize}

        \noindent In particular, we explored whether the following \color{black} effects would appear:
        \textbf{(1) Would the workshop decrease the number of cognitive bias occurrences? (2) Would the workshop increase the participants' use of debiasing techniques?}

            
\color{black}
        
        To answer this research question, we performed 18 controlled experiments with 16 students and 20 software practitioners, divided into control and workshop groups. 

        \noindent
            The study's main \textbf{contribution} is the design and empirical evaluation of the workshop, which resulted in the following:
            \begin{enumerate}
                \item \color{black}The workshop reduced the occurrence of biases for both students and practitioners. It decreased \color{black} the occurrence of all three researched biases (\textit{anchoring}, \textit{confirmation bias}, \textit{and optimism bias}). The decrease in anchoring and optimism bias was statistically significant. 
                \color{black}
                \item The workshop increased the use of all three proposed debiasing techniques. This change was statistically significant in the case of the ``listing multiple solutions'' and ``discussing drawbacks'' techniques. 
                
                \item The workshop significantly improved the amount of non-biased arguments for and against architectural solutions while reducing biased arguments.
                \color{black}
                \item The students discussing their university projects were less impacted by cognitive biases than practitioners in the case of real-life architectures.
                \item The positive effects of the workshop on practitioners were more substantial, 
                \color{black}
                i.e., the decrease in bias occurrences was greater for practitioners. This change was possible because of the higher initial bias level of practitioners, which left more room for improvement.
                \item A set of teaching suggestions based on the factors that impacted the few unsuccessful participants.
                \color{black}
            \end{enumerate}

            Section \ref{sec:related_work} provides information on previous research relevant to this study. Section \ref{sec:Method} describes the research method. The results of the experiment are presented in Section \ref{sec:Results} and discussed in Section \ref{sec:Discussion}. Section \ref{sec:ThreatsToValidity} explores the threats to validity.
            Finally, we summarize this study in Section \ref{sec:Conclusion}.
    
    \section{Related work}
    \label{sec:related_work}
    
    \subsection{Cognitive biases}
    The seminal work of Tversky and Kahneman from 1974~\cite{Tversky1974} introduced the concept of cognitive biases, describing three of them: \textit{representatives}, \textit{availability}, and \textit{anchoring}. In their work, the researchers found that human beings rely heavily on heuristics during the decision-making process while often being blind to logical, statistical facts.  

    These findings later evolved into the dual process theory, describing the human mind as divided into Systems 1 and 2. System 1 performs fast and intuitive decisions that heavily rely on heuristics. Inversely, System 2 performs slow decisions that are logical and rule-based. By using energy mainly for important decisions (System 2), this natural phenomenon allows the human body to save precious energy when making simplified decisions (System 1).
    However, humans are prone to using the energy-saving System 1 for decision-making, even in cases that require rule-based thinking. In these cases, cognitive biases might occur \cite{kahneman2011think}. 

    As a counterpoint to biased reasoning, \textbf{rational reasoning} can be described based on the research by William James~\cite{james1890} as having two components: (1) perception of a specific piece of \textbf{factual information}, and (2) a \textbf{logical consequence} of this information \cite{james1890}. 

    \noindent
    Viewing software architecture as a set of design decisions is a well-established concept~\cite{jansen2005software}. It has resulted in a substantial amount of research regarding issues related to ADM, such as documenting architectural decisions~~\cite{vanHeesch2012documentation}, models of architectural decisions~\cite{zimmermann2009managing}, decision-making best practices \cite{Tang2021}, and human aspects of ADM \cite{Tang2017}, particularly cognitive biases \cite{VanVliet2016}.

    The impact of cognitive biases on ADM can have severe consequences, such as designing sub-par solutions \cite{Manjunath2018a} or incurring dangerous architectural technical debt \cite{Borowa2021b}.

    Based on previous research on cognitive biases in ADM \cite{VanVliet2016,Zalewski2017,Borowa2021b}, as well as the most often researched biases in software engineering \cite{Mohanani2018}, this study focuses on the following three cognitive biases: 

    \textbf{\textit{Anchoring bias}} is the decision-maker's preference for initial information/ideas/so\-lu\-tions (which then become an `anchor') \cite{Tversky1974}. 
    \color{black}
    In ADM, for example, architects may anchor on the first solution idea that comes to their minds, and they may refuse to change or adjust this solution, even when it becomes necessary~\cite{Tang2011}.
    \color{black}

    \textbf{\textit{Confirmation bias}} is the tendency to purposefully search and interpret information to verify one's beliefs \cite{nickerson1998confirmation}. 
\color{black}
    For example, due to confirmation bias, an architect may have the belief that microservices are the best architectural style and insist on designing a microservice-based system, even when it is not appropriate~\cite{Zalewski2017}.
    \color{black}
    
    \textbf{\textit{Optimism bias}} is the overestimation of the probability of positive future outcomes~\cite{sharot2011optimism}. 
    \color{black}
    This may affect architects by leading them to underestimate the maximum amount of requests that would be sent to a particular component~\cite{Zalewski2017}.
    \color{black}

\begin{table}
\caption{Participants \\ \tiny{W-workshop group / C-Control group, \\ 
Age - Age in years, \\
IT exp. - commercial experience in years, \\
DOM - Company domain, \\
SIZE - Company size: S = below 100, M=101-500, L=501-5000, XL= over 5000
}}
\label{tab:participants}
\center
\footnotesize
\scalebox{0.8}{
\begin{tabular}{|c|c|c|c|c|c|c|c|}
\hline
&No&Pair & C/W   & IT   & Role & DOM & SIZE \\
&   &  &      & exp.&       &   &       \\
\hline

\multirow{4}{*}{\rotatebox[origin=c]{90}{Pilot}}
&1& P1 &W   & 6-10  &	Developer				& Digital					& S  \\
& &   &  &   &    			& payment					&   \\
&2& P1 &C   & 1-2  & 	Developer				& Digital					& S\\%
& &   &  &   &    		& payment					&   \\
&3 & P2 &W  & 11-20 &  Analyst & Digital & S \\
& &   &  &   &    	& payment					&   \\
&4& P2 &C  & 3-5 &    Product & Digital & S \\
&&&&  &	Manager			& payment 				&  \\
\hline 
\multirow{14}{*}{\rotatebox[origin=c]{90}{Main - practitioners}}
&5&P3 &W & 11-20 &	Developer		& Media						& XL \\
&6&P3&C	& \textgreater 20 &	Developer		& Media						& XL \\
&7&P4&W	& 11-20 &	Developer		& Marketing& M \\
&8&P4&C	& 11-20 &	 Developer		& Marketing& M \\
&9&P5&W	& 11-20 &	CTO								& Finance					& M \\
&10&P5&C& \textgreater 20 &	Architect						& Finance					& M \\
&11&P6&W	& 6-10	 &	Architect 	& Audio 		& L \\
&12&P6&C& \textgreater 20 &	Sys. Eng.				& Audio	& L \\
&13&P7&W	& 11-20 &	Architect				& Retail					& XL \\
&14&P7&C & 11-20 &	Architect				& Retail					& XL \\

&15&P8&W&  3-5	 &	Developer 	& Education		& L \\
&16&P8&C& 3-5 &	Developer				& Education		& L \\
&17&P9&W& 11-20 &	Product			& Government					& S \\
&&&&  &	Manager			& 				&  \\
&18&P9&C& 11-20 &	Coordinator				& Government					& S \\
&19&P10&W& 1-2 &	Developer				& Education					& L \\
&20&P10&C& 1-2 &	Developer				& Education					& L \\
\hline
\multirow{14}{*}{\rotatebox[origin=c]{90}{Main - students}}
&	21	&	S1	&	W	&	-	&	-	&	-	&	-	\\
&	22	&	S1	&	C	&	1-2	&	Developer	&	Mobile	&	XL	\\ 
&	23	&	S2	&	W	&	\textless 1	&	Developer	&	Electronics	&	XL	\\ 
&	24	&	S2	&	C		&	3-5	&	Developer	&	Utilities	&	L	\\ 
&	25	&	S3	&	W	&	1-2		&	Developer	&	Insurance&	S	\\ 
&	26	&	S3	&	C	&	1-2		&	Developer	&	ERP	&	M	\\ 
&	27	&	S4	&	W	&	\textless 1		&	Developer	&	Ecommerce	&	S	\\ 
&	28	&	S4	&	C	&	1-2		&	Developer	&	Security	&	XL\\ 
&	29	&	S5	&	W	&	-	&	-	&	-	&	-	\\
&	30	&	S5	&	C	&	-	&	-	&	-	&	-	\\
&	31	&	S6	&	W	&	1-2	&	Tester	&	\mbox{Web dev.}&	S	\\ 
&	32	&	S6	&	C	&	1-2		&	Developer	&	ERP	&	M	\\
&	33	&	S7	&	W	&	1-2		&	Pre-sales	&	Sales	&	XL	\\
&	34	&	S7	&	C	&	\textless 1		&	Developer&	Energetics 	&	L	\\
&	35	&	S9	&	W	&	3-5	&	Developer	&	Security	&	L	\\
&	36	&	S8	&	C	&	1-2	&	DevOps	&	Earth	&	M	\\
&		&		&		&		&		&	Observation	&	\\
\hline
\end{tabular}
}
\vspace*{-1.5\baselineskip}
\end{table}

    \subsection{Debiasing} 
Debiasing refers to the process of identifying and mitigating cognitive biases that may affect decision-making.
    Possible types of debiasing treatments can be described using the levels proposed by Fischhoff \cite{fischhoff1982debiasing}. A- and B-level treatments consist of informing practitioners about cognitive biases (A) and how they may impact practitioners (B). Levels C and D require more organizational resources: a C-level treatment requires giving personalized feedback to each debiased person, and the D-level requires extensive long-term training.

    In the field of software engineering, two notable C-level debiasing treatments were reportedly successful. Firstly,  rationalizing development time estimates by Shepperd et al.~\cite{Shepperd2018}, where researchers purposefully anchored software developers on pessimistic estimates to counter over-optimistic predictions.     
    Secondly, in ADM, in their short paper, Borowa et al. \cite{Borowa2022} reported achieving a debiasing effect on a group of student participants. However, their experiment had several weaknesses that our study avoids:
    \begin{enumerate}
        \item The experiment employed the pretest-posttest design, which is known for numerous threats to validity~\cite{Knapp2016}. For example, the debiasing effect may have not resulted from the workshop but from the student participants' learning while performing the pretest task.
        \item The participants were all students\color{black}, making it unclear whether the results would be the same for \color{black} experienced practitioners.
        \item The participants performed different tasks before and after the workshop, so comparing these may not have been reliable. 
        \color{black}
        For example, if designing one architecture was more difficult than the other, participants' bias occurrence count could be increased by this added difficulty. 
        \color{black}
        \item The participants performed purely theoretical tasks by designing architectures for domains that they may not have had any prior knowledge about. This may have caused mistakes due to their lack of domain knowledge.
        \color{black}
        \item The debiasing effect was achieved by improving the amount of non-biased arguments, \textbf{but, despite the authors' efforts, the bias occurrence was not reduced}. We measured the bias occurrences using a similar method to Borowa et al.~\cite{Borowa2022} and did achieve a bias reduction. 
        \color{black}
    \end{enumerate}

    

    \section{Method}
    \label{sec:Method}
    We ran an experiment built on top of an existing debiasing workshop structure~\cite{Borowa2022}.
    However, we improved the workshop and experiment, as explained in the following subsections.
    
    The study was divided into two phases: (1) a pilot phase with two experiments, i.e., two pairs of practitioners from one company, and (2) the main experiment, i.e., with eight pairs of students and eight pairs of practitioners from seven different companies.
    While we present the pilot's results in this paper, the study plan underwent major changes between these phases, which led us to exclude the pilot's data from quantitative calculations (i.e., code count averages and p-values).
    The differences between the pilot and the main study are also explained in the respective subsections.

\begin{figure*}[h]
\centering
\includegraphics[width=0.8\textwidth]{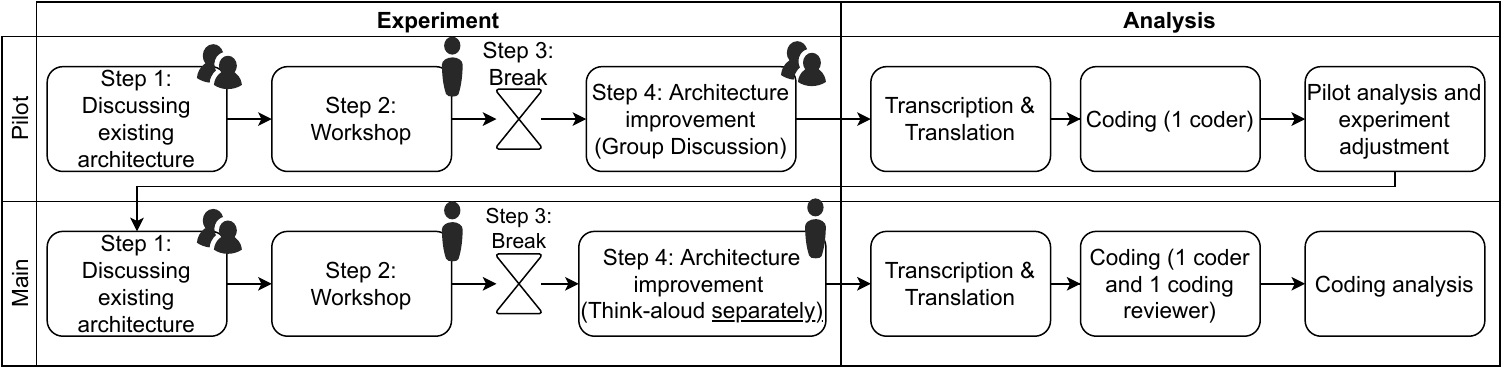}
\caption{Study steps}
\label{fig:experiment_steps}
\end{figure*}
  
\subsection{Sample}
\label{sec:sample}
We recruited two types of participants: students and practitioners. 
\color{black}
We define students as individuals studying to obtain a computer science-related degree who had performed major ADM tasks in an industrial setting. Practitioners, on the other hand, are individuals who have performed major ADM tasks in an industrial setting.
\color{black}
Student pairs discussed projects from their coursework during this study, while practitioners discussed real-life industrial projects. However, it is worth noting that most student participants were also young practitioners with some industrial experience. 

The first author recruited student participants from Warsaw University of Technology through convenience sampling, by advertising the search for participants through their institution's channels. 
To qualify for participation, a pair of students had to have developed a software system from scratch as part of a team during their coursework, i.e., they had to have taken part in requirements gathering, architectural design, implementation, and testing of a system with at least three major components. Sixteen students applied and met these criteria.
\color{black}
Students could obtain extra credit in a software architecture course by participating, though only two students used this opportunity. 
\color{black}

We recruited practitioner participants through convenience sampling from our network in three countries (Brazil, Germany, and Poland) \color{black} and offered no rewards to practitioner participants. \color{black} 
This was the most appropriate method of acquiring participants for two reasons:
\begin{itemize}
    \item The experiment required participants to discuss existing systems used by their companies. As such, participants needed to trust the researchers to handle sensitive data appropriately, which allowed us to face real-world designs instead of theoretical examples.
    \item 
    We needed to find pairs of practitioners who had worked together on developing the same system and who had similar roles. Explaining the nuances to the participants required personal discussion with one of the researchers.
\end{itemize}

The participants were informed beforehand that the experiment was related to improving ADM and about the overall steps of the experiment. However, they were not informed \color{black} about cognitive biases, had no access to the workshop materials, and did not know the Step 4 task of the experiment. No pair of participants had the chance to contact another pair to disclose such information. 
\color{black}
Table \ref{tab:participants} presents basic information about the participants. 

\subsection{The experiment}
\label{sec:experiment}
\color{black}
We conducted an experiment based on an existing debiasing workshop structure~\cite{Borowa2022} with several modifications.
\color{black}
\color{black}
First, we replaced a debiasing practice that was having one person monitor the discussion for statements influenced by \textit{confirmation-bias}, since the previous workshop's authors admit that their participants were rarely able to use this technique at all~\cite{Borowa2022}. 
\color{black}
Instead, our participants were taught to list multiple solution options for their architectural decisions.

\color{black}
Second, 
       we performed an experiment with a matched pair design~\cite{myers2006experimental} instead of the pretest-posttest design. 
       As such, we compared the results for the same architectural task done by matched pairs: (1) where one had attended the workshop (workshop group), and (2) the other had not (control group). 
By \color{black} contrast, in the experiment of~\cite{Borowa2022}, researchers compared different tasks done before and after the workshop by the same students. 
\color{black}
The matched pair design, combined with recruiting pairs of individuals who had worked on a project together, allowed us to explore \textbf{whether the discussed software systems could have benefited from having an individual participate in the debiasing workshop,} while enhancing internal validity by randomizing each pair's split into workshop/control groups.

Third, the architectural designs used in the experiment came from students' coursework projects or real-life cases from practitioners' companies, not hypothetical scenarios.
\color{black}

The experiment itself consisted of (see Figure~\ref{fig:experiment_steps}):
\begin{enumerate}
    \item \textit{Step 1: Discussing existing architecture (up to 30 min.):} Both participants took part in a meeting with a researcher. First, they were asked to fill out a questionnaire to gather demographic data. 
    Then, they were asked to explain the architecture of a system they had both worked on previously. In the case of student participants, they had to discuss a university project where they performed the whole development process, from gathering requirements through architectural design, implementation, and testing. Practitioner participants discussed real-life industrial projects.    
    Participants drew a simple ``boxes and arrows'' diagram of the architecture and gave the researcher the following information: (1) the overall idea and context of the system, (2) the system's main components, (3) relationships between components, and (4) the technologies used. This way, we ensured that both participants knew the architecture at the same level.
    
    \item \textit{Step 2: Debiasing workshop (around 60 min.):} During this step, the workshop group participant took part in the debiasing workshop, as described in Section \ref{sec:workshop}. The control group participant had a break.
    
    \item \textit{Step 3: Break:} 
    Research on cognitive biases~\cite{kahneman2011think} notes that rational logic-based thinking is physically exhausting. To avoid tiring the workshop participants more than the control participants, there was always a break before the last step of the experiment. The length varied but, was always long enough for at least one meal. 
    
    \item \textit{Step 4: Architecture improvement (up to 30 min.):} \color{black} Participants were asked to identify issues with the Step~1 system's architecture. They were asked to propose possible improvements to resolve these issues. \color{black}  The participants were requested to note the changes on the boxes-and-arrows diagram from Step~1. 
    
    \textbf{In the pilot run} with two pairs, we allowed participants to discuss together. 
    However, in one pilot experiment, we noticed that this approach caused a participant to omit their opinions. This was caused by the fact that the other participant, their superior, dominated the discussion. 
    
    \textbf{In the main experiment}, we changed our data gathering approach to a ``think-aloud protocol'' session \cite{Ericsson1993}. 
    This method was suggested by Razavian et al. \cite{Razavian2019} for studying behavioral factors in ADM. Accordingly, the participants proposed their improvements separately     \color{black} 
    (i.e. one after another, in a separate room, with no option to hear each other) in the subsequent sessions, to avoid interferences between participants.
    The participants \color{black} were asked to voice their thoughts, i.e., ``think aloud'', during the whole session. We could not use evaluation methods such as questionnaires or having the participant write down their argumentation since cognitive biases occur during an individual's thought process, and their occurrence may not be visible when the participants have additional time to curate their answers. 
    
\end{enumerate}

The experiment's Step 4 was recorded, with the audio recordings transcribed for further analysis. All the experiment sessions were performed in the native language of the participants by a native-speaking researcher, who subsequently translated the transcript to English to make it accessible to all the authors. We allowed participants to choose between on-site and online participation. Four experiments were performed during a personal meeting (two pilot and two practitioner main), and the rest were held online (all main). 
\color{black}
All the experiments for each pair were performed separately, so no step was performed with many pairs present in the same room at the same time. 
\color{black}
The experiment plan is available as part of the additional material~\cite{additional_material}.


\begin{table*}
\caption{Coding scheme: adapted from \cite{Borowa2022}}
\label{tab:codes}
\scalebox{0.8}{
\begin{tabular}{|p{.05\textwidth}|p{.15\textwidth}|p{.9\textwidth}|}
\hline
Code & Code meaning & Description\\
\hline
\textbf{Arg}  & \textbf{Argument} & A statement in \textbf{support} of a possible solution alternative. \color{black}E.g.~``MS SQL would be a better choice than Oracle since it is the cheaper solution.''\color{black} \\
\textbf{Carg} & \textbf{Counterargument} & A statement in \textbf{opposition} of a possible solution alternative. \color{black}E.g.~``This solution requires turning off the access to users during maintenance''.\color{black}\\
\textbf{Anch}	& \textbf{Anchoring}	& A statement suggesting that the participant is impacted by \textbf{anchoring}. \color{black}E.g.~``Django would be the best choice'' when no additional contextual information suggests why Django was chosen or whether alternatives were considered.\color{black}\\
\textbf{Conf}	& \textbf{Confirmation bias}	& A statement suggesting that the participant is impacted by \textbf{confirmation bias}. \color{black}E.g.~``We already decided on a centralized system, so let's leave it like this,'' when issues with this choice were being discussed. \color{black}\\
\textbf{Opt}	& \textbf{Optimism bias}	& A statement suggesting that the participant is impacted by \textbf{optimism bias}. \color{black}E.g.~``I am sure we can easily connect to some external REST API to handle the payment,'' without researching or having prior knowledge about payment systems. \color{black}\\
\textbf{Ddraw}	& \textbf{Decision's drawback}  & Use of the \textbf{anti-anchoring technique}, i.e., a statement where the participant discusses a drawback of the solution alternative. \color{black}E.g.~``Splitting this service into microservices would require additional work on orchestration.''\color{black}\\ 
\textbf{Dmulti}	& \textbf{Decision with multiple alternatives} & Use of the \textbf{anti-confirmation bias and anti-anchoring technique}, i.e., a statement where the participant mentions more than one solution alternative. \color{black}E.g.~``We could consider various free RDBMS: like MySQL, PostgreSQL, Maria DB or even SQLite.''\color{black}\\
\textbf{Drisk}	& \textbf{Decision's risk} & Use of the \textbf{anti-optimism bias technique}, i.e. a statement where the participant discusses a risk associated with a solution alternative. \color{black}E.g.~``This SSH Server is a possible bottleneck in the system and may stop responding due to too many requests.''\color{black}\\ 
\hline
\end{tabular}
 }
\end{table*}

\subsection{Debiasing Workshop}
\label{sec:workshop}

Our workshop was an intervention that included not only informing about cognitive biases and the specific effects that they may have on the participants but also trained the participants through a teaching experience with personalized feedback. 
Therefore, it is a C-level debiasing intervention on Fischoff’s debiasing scale \cite{fischhoff1982debiasing}. 

Overall, the workshop comprised (1) a short lecture about cognitive biases and the work of Daniel Kahneman and Amos Tversky in general~\cite{Tversky1974}~\cite{kahneman2011think}, (2) an explanation of how cognitive biases affect ADM in particular, (3)
a design session, during which the participant was supposed to design a solution for a fictional architecture task and make use of the debiasing techniques. 
We taught the following debiasing techniques to the participants:
\begin{enumerate}
\color{black}
\item  \textit{Generating multiple solution options:} This is a new debiasing technique that aims to counter \textit{anchoring} and \textit{confirmation biases}.
 The main impact of both these biases is limiting the scope of considered solution alternatives~\cite{Zalewski2017}, i.e., anchoring makes individuals fixated on one solution idea~\cite{Tang2011} and confirmation bias makes them unlikely to change their initial decision based on evidence~\cite{borowa2021knowledge}. As such, it seems prudent to explicitly list solution alternatives before architects believe that one single solution is the best option before considering any others.

Listing multiple solutions is also a basic reasoning technique suggested by many ADM researchers \cite{Tang2021}~\cite{soliman2021exploring}~\cite{cervantes2016designing} for improved reasoning.
\color{black}



\item \textit{Listing at least one drawback of the solution alternative:} This technique is targeted towards countering the effects of \textit{anchoring}, since \textit{anchoring} often makes practitioners over-focused on one solution's advantages \cite{Borowa2022}. Unlike risks, drawbacks always impact the solution, e.g., a component's license is expensive. 
\item \textit{Listing at least one risk associated with the solution alternative:} Since individuals are prone to over-optimistic predictions, explicitly discussing risks is targeted towards countering \textit{optimism bias} \cite{Borowa2022}.
Risks, unlike drawbacks, have a probability associated with their occurrence. This means there is a possibility that the risk may not occur at all.
\end{enumerate}

The researcher leading the workshop demonstrated the debiasing techniques using examples and actively assisted the participants in the design session when they were struggling with the techniques. 


\subsection{Data Analysis}
\label{sec:data_analysis}

For the data analysis procedure, we used the hypothesis coding technique \cite{Saldana2013}, which involved creating a set of codes before the coding process. 
The codes that we used are listed in Table \ref{tab:codes}
\color{black}
and were prepared based on the same coding scheme used by Borowa et al.~\cite{Borowa2022}. 
\color{black} 
A coding guide with examples from this study's participants is available online~\cite{additional_material}.

Recordings from the sessions were transcribed and translated to English by the researcher organizing the workshop, then coded by another author and, for the main experiment, the coding was reviewed by an additional author.

In the pilot phase, the coder was unaware of which \color{black}participant was the control/workshop one. 
The main experiment, as suggested by Ericsson and Simon~\cite{Ericsson1993}, \color{black} employed \textbf{context-free coding} as follows:
\begin{enumerate}
    \item The workshop organizer divided the transcripts into segments (one for each architectural decision). This was a natural split since the participants usually discussed one possible change to the architecture after another.
    \item For each pair of participants, the control and workshop groups' segments were randomly ordered. 
    \item Another researcher coded the segments, without knowing whether a particular segment belonged to the workshop or control group participant. 
    \item An additional researcher reviewed the coding and discussed change proposals with the original coder until a consensus was reached on all codes.
    \item The workshop organizer revealed the participants' information to create a summary of each experiment's coding.
\end{enumerate}

      \begin{figure*}[!htb]
            \centering
            \includegraphics[width=0.85\textwidth]{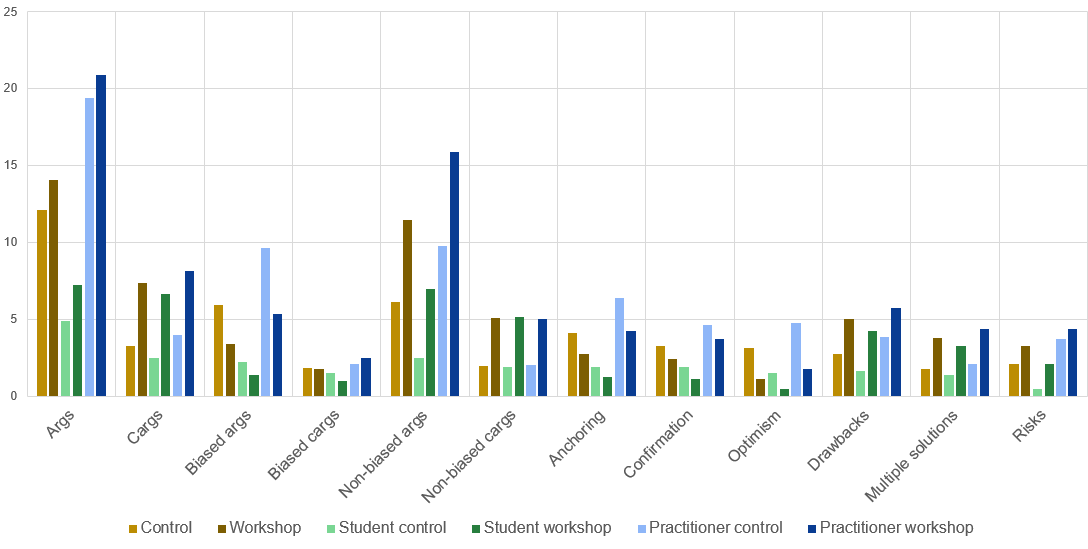}
            \caption{Average amount of codes for control and workshop group for (1) All participants, (2) students, (3) practitioners.}
            \label{fig:codes_avg}
        \end{figure*}
        
\begin{table*}[!htb]
\centering
\caption{(Non-)Biased Arguments and Counterarguments \tiny{(B -- Biased, NB -- Non-biased)}}
\label{tab:arguments}
\scalebox{0.9}{
\begin{tabular}{{|c|c|p{0.7cm}|p{0.7cm}|p{0.7cm}|p{0.7cm}|p{0.7cm}|p{0.7cm}|p{0.7cm}|p{0.7cm}|p{1.7cm}|p{1.7cm}|p{1.7cm}|}}
\hline
& & \multicolumn{4}{c|}{Control}   & \multicolumn{4}{c|}{Workshop}  & \textbf{Control} & \textbf{Workshop} & \textbf{Difference}\\
& & \multicolumn{2}{c|}{Arguments}& \multicolumn{2}{c|}{Counterarg.} &\multicolumn{2}{c|}{Arguments}& \multicolumn{2}{c|}{Counterarg.} &\textbf{Biased} &\textbf{Biased } &\textbf{Biased }\\
 &Pair& B  &NB  &B &NB &B &NB &B&NB &\textbf{statements [\%]} &\textbf{statements [\%]} &\textbf{statements [pp]}\\
\hline

\multirow{2}{*}{\rotatebox[origin=c]{90}{Pilot}}
	&	P1	&	5	&	2	&	1	&	2	&	4	&	0	&	1	&	6	&	60\%	&	38\%	&	-22	\\
	&	P2	&	8	&	1	&	0	&	1	&	1	&	3	&	0	&	1	&	80\%	&	17\%	&	-63	\\
\hline
\multirow{7}{*}{\rotatebox[origin=c]{90}{Main: practitioner}}
	&	P3	&	12	&	10	&	4	&	1	&	8	&	8	&	2	&	9	&	59\%	&	36\%	&	-24	\\
	&	P4	&	3	&	6	&	1	&	1	&	11	&	5	&	9	&	4	&	36\%	&	67\%	&	30	\\
	&	P5	&	14	&	22	&	1	&	4	&	3	&	20	&	0	&	1	&	37\%	&	11\%	&	-26	\\
	&	P6	&	9	&	10	&	2	&	3	&	5	&	13	&	3	&	7	&	46\%	&	26\%	&	-20	\\
	&	P7	&	24	&	17	&	6	&	4	&	10	&	54	&	3	&	13	&	59\%	&	15\%	&	-43	\\
	&	P8	&	3	&	1	&	0	&	1	&	1	&	10	&	0	&	1	&	60\%	&	8\%	&	-52	\\
	&	P9	&	7	&	9	&	2	&	1	&	3	&	11	&	2	&	4	&	47\%	&	24\%	&	-24	\\
&	P10	&	5	&	3	&	1	&	1	&	2	&	6	&	1	&	1	&	60	\%&	30	\%	&	-30	\\
\hline
\multirow{7}{*}{\rotatebox[origin=c]{90}{Main: student}}
&	S1	&	1	&	3	&	1	&	1	&	1	&	4	&	0	&	8	&	33	\%&	8	\%	&	-25	\\
&	S2	&	1	&	3	&	1	&	1	&	1	&	8	&	0	&	1	&	33	\%&	10	\%	&	-23	\\
&	S3	&	1	&	2	&	2	&	1	&	1	&	15	&	2	&	11	&	50	\%&	10	\%	&	-40	\\
&	S4	&	2	&	1	&	2	&	0	&	1	&	5	&	1	&	3	&	80	\%&	20	\%	&	-60	\\
&	S5	&	3	&	1	&	3	&	0	&	3	&	3	&	3	&	5	&	86	\%&	43	\%	&	-43	\\
&	S6	&	2	&	1	&	0	&	5	&	4	&	4	&	1	&	2	&	25	\%&	45	\%	&	20	\\
&	S7	&	5	&	7	&	2	&	5	&	0	&	11	&	0	&	4	&	37	\%&	0	\%	&	-37	\\
&	S8	&	3	&	2	&	1	&	2	&	0	&	6	&	1	&	7	&	50	\%&	7	\%	&	-43	\\

\hline
\end{tabular}
}

\vspace*{-1.5\baselineskip}
\end{table*}

\begin{table*}[!htb]
\centering
\caption{Occurrences of Cognitive Biases \tiny{(ANCH -- Anchoring, OPT -- Optimism bias, CONF -- Confirmation bias)}}
\label{tab:biases}
\scalebox{0.9}{
 \begin{tabular}{|c|c|r|r|r|r|r|r|r|r|r|} 
\hline
& & \multicolumn{3}{c|}{Control}   & \multicolumn{3}{c|}{Workshop}  & \textbf{Control} & \textbf{Workshop} & \textbf{Difference}\\
 & Pair &ANCH &CONF  &OPT &ANCH &CONF &OPT &\textbf{Bias sum} &\textbf{Bias sum} &\textbf{Bias sum}\\
\hline
\multirow{2}{*}{\rotatebox[origin=c]{90}{Pilot}}
	&	P1	&	6	&	3	&	1	&	1	&	1	&	1	&	10	&	3	&	-7	\\
	&	P2	&	7	&	3	&	1	&	1	&	0	&	0	&	11	&	1	&	-10	\\
\hline
\multirow{7}{*}{\rotatebox[origin=c]{90}{Main: practitioner}}
	&	P3	&	9	&	7	&	2	&	7	&	4	&	0	&	18	&	11	&	-7	\\
	&	P4	&	3	&	1	&	2	&	12	&	11	&	0	&	6	&	23	&	17	\\
	&	P5	&	10	&	3	&	8	&	1	&	2	&	3	&	21	&	6	&	-15	\\
	&	P6	&	9	&	3	&	0	&	5	&	3	&	1	&	12	&	9	&	-3	\\
	&	P7	&	6	&	21	&	23	&	6	&	6	&	4	&	50	&	16	&	-34	\\
	&	P8	&	1	&	0	&	3	&	0	&	0	&	2	&	4	&	2	&	-2	\\
	&	P9	&	7	&	2	&	0	&	2	&	4	&	2	&	9	&	8	&	-1	\\
 &	P10	&	6	&	0	&	0	&	1	&	0	&	2	&	6	&	3	&	-3	\\
 \hline
\multirow{7}{*}{\rotatebox[origin=c]{90}{Main: student}}
&	S1	&	0	&	1	&	1	&	0	&	1	&	1	&	2	&	2	&	0	\\
&	S2	&	0	&	1	&	1	&	1	&	0	&	0	&	2	&	1	&	-1	\\
&	S3	&	3	&	2	&	3	&	1	&	1	&	1	&	8	&	3	&	-5	\\
&	S4	&	2	&	1	&	1	&	1	&	1	&	1	&	4	&	3	&	-1	\\
&	S5	&	4	&	3	&	1	&	4	&	3	&	1	&	8	&	8	&	0	\\
&	S6	&	2	&	1	&	1	&	2	&	3	&	0	&	4	&	5	&	1	\\
&	S7	&	4	&	4	&	0	&	0	&	0	&	0	&	8	&	0	&	-8	\\
&	S8	&	0	&	2	&	4	&	1	&	0	&	0	&	6	&	1	&	-5	\\

\hline
\end{tabular}
}
\vspace*{-1.5\baselineskip}
\end{table*}

\color{black}We anticipated that a coder might unconsciously try to find codes proving the research hypotheses.
The context-free coding method aims to decrease the possibility of this occurring. \color{black}

We evaluated our coding as follows: (1)~We calculated the average of all codes for a particular measurement for each group of participants. (2)~We checked whether the measurements for cognitive biases and bias-impacted statements decreased. (3)~We checked whether the measurements of debiasing techniques and non-biased statements increased. 

In the work of Borowa et al. \cite{Borowa2022}, the authors assumed that if most arguments used while discussing one decision were non-biased, then the decision was non-biased. 
We have not followed this assumption since one argument (biased or non-biased) can be decisive in decision-making. 

In order to determine whether the \color{black} debiasing intervention \color{black} was successful, we defined a set of research hypotheses (H\textsubscript{R}):
\begin{itemize}
    \item In the case of \textbf{values that the workshop strived to decrease} (biased arguments/counterarguments, cognitive biases), the research hypothesis (H\textsubscript{R}) was that the measurements with the workshop (workshop group) were smaller than without it (control group). Our null hypothesis (H\textsubscript{0}) was that the control group measurements were smaller than or equal to the workshop group measurements. 
    \item The opposite was the case with \textbf{values that the workshop tried to increase} (not biased arguments/counter\-ar\-gu\-ments, use of debiasing techniques). \color{black} In this case, \color{black} our research hypothesis (H\textsubscript{R}) was that the measurements with the workshop (workshop group) were bigger than without it (control group). In these cases, the null hypothesis (H\textsubscript{0}) was that the control group measurements were bigger (or equal to) the workshop group measurements. 
\end{itemize}

To examine which results are statistically significant and can be considered as accepted with a high confidence level, we performed the non-parametric Wilcoxon Signed Rank test \cite{wilcoxon1992individual} for each of the measurements and the percentage of biased statements (i.e. arguments and counterarguments). This test is used to measure whether a statistical difference can be observed between small paired (dependent) data samples. We consider the control group and workshop group participant pairs to be dependent since they both shared an understanding of the particular architecture and performed the task of discussing improvements for the same systems.
We accept H\textsubscript{R} as statistically significant if the p-value is smaller than 0.05, i.e., there is a less than 5\% chance that H\textsubscript{R} is false.



\section{Results}
    \label{sec:Results}

In this section, we present the results of our analysis. Due to the change in data-gathering methods, the results from the pilot study were not included in any calculations (p-values, overall code sum/average for the control/workshop group), but raw values are presented in this section as well.

Figure \ref{fig:codes_avg} showcases the average code counts for each measurement totaled across all participants, \color{black}i.e. for each code, we summed the number of occurrences in the transcript for each participant and calculated the averages for participant groups. \color{black}When considering only the code averages, all measured values changed in accordance with the research hypothesis H\textsubscript{R}. That is: (1) the amount of arguments and counterarguments increased, (2) the amount of biased arguments and counterarguments decreased, (3) the amount of non-biased arguments and counterarguments increased, (4) the amount of bias occurrences decreased, (5) the amount of debiasing techniques use increased.

\subsection{\color{black}Statistical Significance\color{black}}

        \begin{table}[!htb]
\caption{p-values for each measurement \\ \tiny{W-workshop group / C-Control group, \\
AVG - Average of code sums, \\
H\textsubscript{R} - Research hypothesis\\
}}
\label{tab:pvalues}
\center
\scalebox{0.9}{
\begin{tabular}{{|p{2.5cm}|p{1cm}|p{1cm}|p{1cm}|p{0.8cm}|}}
\hline
Measurement	&	C-AVG & W-AVG & p-value	& (H\textsubscript{R})
 \\
\hline
Arg								& 12.13 		& 14.06      	&	0.1619 				&	W \textgreater C	\\
\textbf{Carg	}			            & 3.25 		& 7.28      	& \textbf{0.0022	}	&	W \textgreater C	\\
\hline           			        	           	
\textbf{Anch	}						& 4.13 		& 2.75       	&	\textbf{0.0414	}&	W \textless C	\\
Conf							& 3.25 		& 2.24       	&	0.2062				&	W \textless C	\\
\textbf{Opt	}				            & 3.13 		& 1.13        	&	\textbf{0.0404	}&	W \textless C	\\
\textbf{Biases sum	}					& 10.50 		& 6.31      	& \textbf{0.0108	}			&	W \textless C	\\
\hline          			        	           	
\textbf{Biased Args	}					& 5.94 		& 3.38       	&	\textbf{0.0205}			&	W \textless C	\\
Biased Cargs					&  1.81		& 1.71       	&	0.1482					&	W \textless C	\\
\textbf{Non-biased Args	}				& 6.13 		& 11.44 		& \textbf{0.0016	}		&	W \textgreater C	\\
\textbf{Non-biased Cargs}		& 1.94 		& 5.06      	&	\textbf{0.0057}	&	W \textgreater C	\\
\textbf{\% Biased statements}			& 49,90\% 	& 23,13\% 	&		\textbf{0.0008}			&	W \textless C	\\
\hline                  	            
\textbf{Ddraw	}						& 2.75 		& 5.00      	&	\textbf{0.0123}	&	W \textgreater C	\\
\textbf{Dmulti	}				& 1.75 		& 3.81      	& \textbf{0.0018}		&	W \textgreater C	\\
Drisk							& 2.13 		& 3.25      	&	0.0692				&	W \textgreater C	\\
\textbf{Techniques use sum	}			& 10.57 		& 14.57      	&	\textbf{0.0038	}			&	W \textgreater C	\\

\hline

\end{tabular}
}
\vspace*{-0.5\baselineskip}
\end{table}

The p-values calculated using the Wilcoxon Signed Rank test \cite{wilcoxon1992individual} are presented in Table~\ref{tab:pvalues}. We calculated the p-values for all measurements that we coded and the percentage of biased statements. Most observed measurement changes were statistically significant, with the exception of (1) increased argument amount, (2) decreased amount of biased counterarguments, (3) decreased amount of confirmation bias occurrences, and (4) increased amount of the ``listing risks'' technique uses.

\subsection{Arguments and Counterarguments} 
\label{sec:Arguments}


\color{black}We coded as arguments the statements that supported choosing a particular architectural solution, e.g. Participant 6 said ``(...) what is special about this architecture is that everything is dynamically configurable.''
Counterarguments are statements that give reasons against choosing a solution, e.g. Participant 2 notes price ``(...) AWS itself is another third party service that is paid as well.'' \color{black}
Table \ref{tab:arguments} presents the numbers of arguments and counterarguments for each pair of participants. 
When combined with Figure \ref{fig:codes_avg}, showing the average of each type of statement in the control and workshop groups, it allows us to make the following observations:
\begin{itemize}
    \item The number of arguments and counterarguments used was higher for workshop participants.
    \item Workshop participants used fewer biased arguments and counterarguments overall.
    \item Practitioners used more arguments than students in both workshop and control groups. 
    \item The number of practitioners' biased counterarguments slightly increased after the workshop.
    
\end{itemize}

Additionally, based on the data shown in Table~\ref{tab:pvalues}, we can distinguish a set of statistically significant results related to arguments and counterarguments. \textbf{The following differences between participant pairs were statistically significant: (1) increased counterargument amount, (2) increased non-biased statements amount for both arguments and counterarguments, (3) decreased amount of biased arguments.} However, the increase in arguments amount in favor of a solution was not significant, nor was the decrease in biased counterargument amount.  
Additionally, when looking at the individual values per pair, two participants, one student (pair S6), and one practitioner (pair P4), used more biased statements, despite the workshop's intervention.


\subsection{\color{black}Cognitive Biases\color{black}}
\label{sec:CognitiveBiases}

\color{black}We coded biased statements using three codes. We identified anchoring in arguments favoring solutions without factual data or solution alternatives, e.g., Participant 14 said the following, without providing additional rationale: ``I would move all this work on data, manipulation of this data, certainly, to the server.'' 
Second, we used the confirmation bias code when participants avoided discussing improvements despite identifying problems; Participant 7 remarked: ``Even if we were to talk about splitting up the middleware or converting it to microservices or other structures, I wouldn't do any of that because that's what [colleague] and I have defined as the actual system.''
Finally, optimism bias was coded for overly optimistic predictions, e.g., Participant 36 proposed parallel processing, stating ``Surely this would also speed up the whole process,'' without considering communication overheads. 
\color{black}

Table \ref{tab:biases} presents the number of biased statements from each participant pair. Figure \ref{fig:codes_avg} summarizes of average bias occurrences in each of the control and workshop groups.
Overall, the workshop group participants were less impacted by each of the researched biases. \textbf{The results were statistically significant (see Table~\ref{tab:pvalues}) in the cases of decreasing anchoring and optimism bias occurrences.}  

However, two differences between practitioners and students are noticeable: \textbf{(1) the average bias occurrence amount (both control and workshop) was higher in the case of practitioners}, and \textbf{(2) decreased bias occurrence was higher in practitioners than students}. Overall, the occurrence of biases of three student pairs (S1, S2, S5) did not decrease, which was the case for only one practitioner pair (P4).






\subsection{\color{black}Debiasing Techniques\color{black}}
\label{sec:DebiasinTechniques}

\color{black}\color{black}

Figure \ref{fig:codes_avg} shows the usage of each technique in the workshop and control groups.
Overall, \textbf{the workshop participants used each of the three techniques more frequently than the control group participants}. This was statistically significant (see Table~\ref{tab:pvalues}) in the cases of ``discussing drawbacks'' and ``listing multiple solutions'', but not for the ``discussing risks'' technique.
Two participants' use of debiasing techniques did not improve after the workshop, one practitioner (pair P5) and one student (pair S6).
\color{black}The code count for each participant's use of debiasing techniques is part of the additional material~\cite{additional_material}.\color{black}


\subsection{Decisions}

\color{black}\color{black}

We did not attempt to influence the decision amount through this experiment, so no p-values were calculated. \color{black}However, during the experiment, we noticed that the amount of discussed decisions varied, so we decided to explore whether decision count could possibly be linked to cognitive bias susceptibility. 
In most cases (11 out of 18), the workshop participants discussed more architectural decisions than their control group collaborators. Pair P4 was an outlier in decision count, with the workshop participant discussing 19 decisions against 9 discussed by the control participant, which was the biggest difference of all the pairs. This pair is the only practitioner pair where the workshop participant's bias occurrences and percentage of biased statements were higher than their control group counterparts. 
Detailed decision count data is presented in the additional material~\cite{additional_material}.
\color{black}

\section{Discussion}
\label{sec:Discussion}
In this section, we discuss our results in the context of (1) comparison with previous work on debiasing ADM, \color{black} (2) differences between student and practitioner participants, (3) a set of teaching suggestions for educators, (4) implications for researchers.\color{black}


\subsection{\color{black}Comparison with Previous Work} 
Compared to the only study \color{black}reporting an effective debiasing ADM intervention by Borowa et al.~\cite{Borowa2022}, \color{black}performed with solely student participants, \textbf{our results show more positive outcomes}. In their study, the amount of non-biased arguments and counterarguments and the use of debiasing techniques were improved. Our study identified similar outcomes and additionally showed:  \textbf{(1) a decrease in bias occurrences (optimism bias and anchoring decreased significantly) and (2) a significant decrease of biased arguments in support of a solution}. This may be the result of two factors: (1) improved experiment and workshop design and (2) the practitioner participants' stronger response to the debiasing workshop \color{black}(i.e. bigger decrease in bias occurrences) \color{black} in comparison to student participants.


\subsection{\color{black}Comparision between Practitioners' and Students'} 
\label{sec:discussion:pract_vs_students}
We discovered that there are notable differences in how students and practitioners reason about architectural decisions and respond to the debiasing workshop.

First, \textbf{practitioners were much more skilled overall in using arguments in support of a solution}, in both control and workshop groups, \color{black} i.e. on average, they used almost twice as many arguments in favor of solutions than students. \color{black} We suspect this may be due to the practitioners' having the preexisting skill of arguing in favor of their chosen solutions since they most likely had to do this repeatedly in their workplace. In contrast, students may find the formulation of any arguments, both for and against architectural solutions, new due to their limited experience. 

Second, \textbf{the number of biased counterarguments slightly increased in the case of practitioners}, which was not the case with students. \color{black} This may be because practitioners had more previous negative experiences (e.g. failed projects), so they had a higher tendency to prematurely dismiss solutions that they had a personal negative experience with.

Thirdly, \textbf{practitioners from both control and workshop groups were overall more impacted by biases than students}\color{black}, i.e., the average bias occurrence measurement for all biases was higher in the case of practitioners. \color{black}As a result, practitioners' decrease in bias occurrence was more extreme than in the case of students. Having industrial experience may actually make practitioners more prone to the impact of cognitive biases. 
This higher bias susceptibility may be due to practitioners' higher attachment to the solutions that they worked on. In the case of a finished student project, the student can freely critique the solution since there is no danger of losing customers or damaging their reputation. Conversely, practitioners spend years polishing their systems, which they have to sell to customers. 
This strong attachment to their solutions is further evidenced by practitioners’ frequent use of arguments to support their choices and the non-significant decrease in the occurrences of confirmation bias—i.e., the tendency to dismiss information that contradicts their beliefs.

\color{black} Similarly, in the software testing domain, Calikli et al.~\cite{calikli2010empirical} found that active software testers were more prone to confirmation bias than ex-testers or trained students. They attributed this to habits formed by routine tasks and suggested breaking routines by periodically performing varying tasks. A similar approach could also benefit software architects, such as participating in non-architectural activities. \color{black}

\subsection{\color{black}Recommendations for Educators}
\color{black}
Overall, the workshop was a successful debiasing intervention and we recommend its use by educators. However, we additionally investigated the participants who did not perform as expected, to explore what educators might take into account to improve this debiasing intervention. 

\textbf{Teaching suggestion 1: Debiasing might only be effective when team members of all seniority levels participate in a debiasing intervention.}
As shown by Participant 3 (pair P2 from the pilot), who discussed only two design decisions, group discussion can be heavily impacted by pressure from senior colleagues. As such, socio-cultural factors associated with group decision-making \cite{muccini2018group} make it crucial to perform the intervention on all individuals, not only a few team members, who may be unable to pass their knowledge about biases to their colleagues. 

\textbf{Teaching suggestion 2: Avoid discussing too many decisions.}
Pair P4 was the only practitioner pair where the workshop participant's bias occurrences were higher than their control group colleague's. The only value that seems to distinguish this pair is that the workshop participant discussed many more decisions than the control group participant. This difference was the highest of all the pairs. As such, discussing too many decisions in a short time span may increase the likelihood of cognitive bias and should be avoided. This might result from a heightened cognitive load~\cite{gonccales2021measuring} experienced by the participant, who attempted to discuss a large number of design decisions swiftly. Psychological research shows that cognitive load impacts \textit{anchored} time estimates~\cite{blankenship2008elaboration}. Thus, we suspect that it may also have impacted our participant. Especially since anchoring was the main bias impacting this participant.

\textbf{Teaching suggestion 3: Address overconfidence to counter susceptibility to cognitive biases.}
Student pair S6 was the only student pair where the workshop group participant did not perform better in almost any regard (bias occurrences, biased statements, and debiasing techniques). This student's case was quite distinct because, during the workshop (before which we did not disclose that the study is about cognitive biases), they informed the researcher conducting the workshop that they had read Kahneman's famous book about cognitive biases~\cite{kahneman2011think}. As such, this participant had a high confidence in their knowledge of cognitive biases.
Similarly, the workshop participant from pair P5 was the only practitioner who used a smaller amount of debiasing techniques than his control group colleague. Notably, the workshop group participant was the company's CTO. During the experiment, the researcher noticed that the attitude of the workshop participants indicated that they considered the workshop's contents trivial. The participant's confidence level was most likely heightened due to their high position in the company, where they were previously the head architect for many years. Therefore, we suggest that educators performing debiasing interventions should warn unusually confident individuals about their possibly higher susceptibility to biases.

\textbf{Teaching suggestion 4: Limit low-quality counterarguments.} In the case of practitioners, a slight increase in the use of biased counterarguments by workshop participants occurred. This may be because the workshop focused on negative factors, such as decision-related risks and drawbacks. As such, some participants may have over-focused on creating numerous counterarguments with less care about their quality. Educators may ask participants to limit the number of counterarguments to a set number of crucial ones during the workshop.

\subsection{\color{black}Implications for Researchers}
Our study provides researchers with a valuable starting point, showcasing how a successful ADM debiasing intervention can be performed. However, we strongly advise that researchers consider the following issues.
First, since confirmation bias was the only cognitive bias that did not decrease significantly, researchers should focus on finding methods of countering its impact. Since we suspect that routine, previous negative experiences and overconfidence may be the source of this problem, debiasing methods could consider routine interruption, discussing previous negative experiences, and warning overconfident individuals.
Second, since practitioners' bias levels were higher and attachment to an existing architectural solution seems to be a strong factor affecting biases, future studies should focus on debiasing practitioners on real-life industrial systems.
Finally, since the use of the ``discussing risks'' technique did not increase significantly, the workshop did not fully teach participants how to perform it. Researchers could attempt to change the workshop's contents related to teaching this technique or even consider replacing it with a different technique altogether.

\color{black}

    \section{Threats to validity}
    \label{sec:ThreatsToValidity}
        Our threats to validity are described based on the guidelines 
        created by Wohlin et al.~\cite{Wohlin2000}:

    	\textbf{Construct Validity --}
     \color{black} It is important to note that every human's thought process is nuanced and multifaceted, so the ``biased statements'' that we measured are a simplification of this process. As such, there exists the threat that our study over-simplified this complex phenomenon by, for example, not taking into account how several biases may impact decision-making simultaneously, and the relationships between them. \color{black}
     
     

         	\textbf{Internal Validity --}
        First, our sample may be impacted by selection bias since our participants were volunteers. It is possible that individuals less interested in architectural decision-making may be impacted by cognitive biases and the workshop differently.
        Second, student participants discussed systems from a university project, while practitioners described real-life systems. Therefore, when comparing their outcomes, it must be noted that some result differences may be rooted in the nature of university projects and not the participant's experience level. We take this into account when discussing our results (see Section~\ref{sec:discussion:pract_vs_students}). This was unavoidable since there was no different way of testing how students reason if they had no experience in designing real-world architectures.
        \color{black}
        Third, while our experiment, which employed a matched pair design, did include randomized splitting into the control/workshop group, internal validity could have been enhanced even further by a fully random assignment design~\cite{ACM_standards_experiment}. However, in that case, all the participants would have had to discuss the same task, making it impossible for them to use their knowledge about the system fully, thus limiting external validity.
        Finally, a key limitation of the experiment lies in the design of the control group. The control group participants were given no alternative intervention and simply took a break. This raises the possibility that the observed improvements may stem from any intervention rather than the debiasing workshop. Future studies should include a placebo~\cite{myers2006experimental} intervention for the control group participants to mitigate this threat. 
        \color{black}

        \textbf{Conclusion Validity --}
         We have the threat of a relatively low number of experiments executed, which is due to the fact that it was hard to find participants, especially practitioners, willing to invest over two hours in a scientific experiment. Finding participants was particularly challenging since our experiment required (1) pairs of participants who worked together on the same project and (2), in the case of practitioners, sensitive data about their real-life architecture.
         This low sample size places a risk on the statistical significance of the findings.
         To counteract this threat, we performed the non-parametric Wilcoxon Signed Rank Test, which makes it possible to check the statistical significance of small paired samples. 
         Still, a larger sample would provide more confidence in the findings.

    \textbf{External Validity --}
    As common for a controlled experiment, the external validity is hampered by the strictly controlled setting. However, we tried to perform the experiment in the participants' preferred environment to lessen this effect. 
    Additionally, the student participants have less domain knowledge than practitioners, which could lead to poor design decisions. We attempted to lessen this effect by having students discuss a project that they designed and implemented themselves during their coursework, which means that they had to analyze the project's domain beforehand.
Finally, we could not research the long-term effect of the debiasing workshop. The workshop's beneficial impact may lessen over time. Future work could focus on studying the long-term effects of our workshop. 

\section{Conclusion}
\label{sec:Conclusion}
We performed a debiasing workshop in order to explore its impact on students and practitioners, i.e. individuals at various career stages.
The design of our study was based on the structure by Borowa et al. \cite{Borowa2022}, with some adaptations (as explained in Section~\ref{sec:Method}). 
We divided participants into separate workshop (treatment) and control groups as it is common in similar studies~\cite{razavian2016two,Tang2017}, striving for a higher level of validity than in a pretest-posttest experiment~\cite{Knapp2016}. We also employed the think-aloud protocol study method suggested by Razavian et al. \cite{Razavian2019} as appropriate for researching behavioral aspects of ADM. 

The research question that we aimed to answer was:
\begin{itemize}
                \item [] \textbf{RQ: How does the proposed debiasing architectural decision-making workshop influence students and practitioners?}
    \end{itemize}
Our main contribution includes the design and evaluation of a debiasing workshop. Through this evaluation, we found:
\begin{enumerate}
    \item The use of the proposed debiasing techniques that were taught increased. In particular, the use of``discussing drawbacks'' and ``listing multiple solutions'' techniques' increased significantly.
     \item The occurrences of cognitive bias for all three researched biases (anchoring, confirmation bias, and optimism bias) decreased. This decrease was significant in the case of anchoring and optimism bias.
    \item Practitioners were commonly more influenced by all researched biases than students, and as such, the improvement achieved in their case was more prominent. 
    \item \color{black} A set of teaching suggestions based on factors that negatively impacted some of the study's participants.\color{black}
\end{enumerate}

\textbf{Practitioners and educators} intending to perform such a debiasing workshop can use the teaching materials we prepared for this study either directly or using them as a basis for their own materials~\cite{additional_material}. \textbf{Researchers} may make use of this study as a basis for creating improved debiasing interventions and evaluate them by employing our experimental design. 

\textbf{For future work,} we propose to focus on (1) longitudinal studies in a real-world setting to ensure that the debiasing treatment has a long-term impact, (2) further improvements meant to reduce the impact of confirmation bias, \color{black}(3) an in-depth qualitative analysis of the biases in ADM.\color{black}

\section{Data availability}
The experiment plan, workshop plan, workshop slides, participant questionnaire, coding results, and a coding guide with examples from the participants are available online~\cite{additional_material}.

\bibliographystyle{IEEEtran}
\bibliography{bibliography}

\end{document}